\documentclass[12pt,prb]{revtex4}%
\usepackage{amsfonts}
\usepackage{amsmath}
\usepackage{amssymb}
\usepackage{graphicx}%
\providecommand{\U}[1]{\protect\rule{.1in}{.1in}}
\begin{document}
\preprint{APS/123-QED}
\title{Magnetization and spin gap in two-dimensional organic ferrimagnet BIPNNBNO}
\author{V.E. Sinitsyn, I.G. Bostrem, A.S. Ovchinnikov}
\affiliation{Department of Physics, Ural State University, 620083, Ekaterinburg, Russia}
\author{Y. Hosokoshi}
\affiliation{Department  of Physical Science, Osaka Prefecture University, Osaka, Japan}
\author{K. Inoue}
\affiliation{Department of Chemistry, Hiroshima University, Hiroshima, Japan}

\date{\today}

\begin{abstract}
A magnetization process in two-dimensional ferrimagnet BIPNNBNO is analyzed. The compound consists of ferrimagnetic (1,1/2) chains coupled by two sorts of antiferromagnetic interactions.  Whereas a behavior of the magnetization curve in higher magnetic fields can be understood within a process for the separate ferrimagnetic chain,  an appearance of the singlet plateau at lower fields is an example of non-Lieb-Mattis type ferrimagnetism. By using the exact diagonalization  technique for a finite clusters of sizes $4 \times 8$ and $4 \times 10$ we show  that the interchain frustration coupling  plays an essential role in stabilization of the singlet phase. These results are complemented by an analysis of four cylindrically coupled ferrimagnetic (1,1/2) chains via an abelian bosonization technique and an effective theory based on the XXZ  spin-1/2  Heisenberg model when the interchain interactions are sufficiently weak/strong, respectively.
\end{abstract}

\pacs{Valid PACS appear here}
\maketitle

\section{Introduction}

During the last fifteen years, two-dimensional (2D) quantum spin systems have attracted
a lot of attention both from theoretical and experimental physicists. A competition between conventional classically  ordered phases and more exotic quantum ordered phases lies in the focus of the investigations. Magnetic systems
with a finite correlation length at zero temperature and a finite spin gap above the singlet ground state,  spin liquids, realize Haldane prediction at the level of two space dimensions \cite{Mila2000}. 

To date, one can distinguish two main routes in studies of 2D spin gap compounds. A formation of spin gap in spin dimer  systems, for example  SrCu${}_2$(BO${}_3$)${}_2$ \cite{Kageyama1999} and CaV${}_4$O${}_9$ \cite{Taniguchi1995}, is explained by a modified exchange topology similar to Shastry-Sutherland lattice \cite{Shastry1981}. Another way to  increase  quantum fluctuations and stabilize a spin liquid ground state is realized in kagome antiferromagnet \cite{Lacroix}.  Experimental candidates for 2D kagome antiferromagnets are currently available:  herbertsmithite\cite{Helton2007,Mendels2007,Lee2007} and volborthite \cite{Bert2005,Yoshida2009}. Both these strategies  deal with antiferromagnetic compounds. In view of this, an observation of  a singlet ground state with a pronounced spin gap in 2D  ferrimagnetic material BIPNNBNO  seems exotic \cite{Goto2003}.

\begin{figure}[ht]
\begin{center}
\includegraphics[width=75mm]{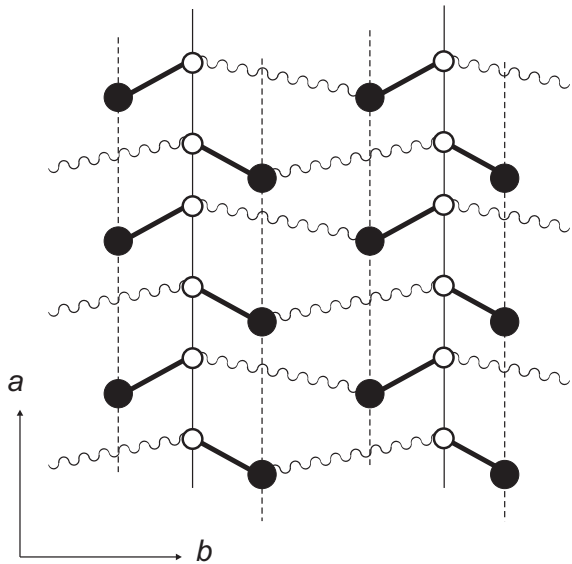}
\end{center}
\caption{Magnetic model of BIPNNBNO crystal. The black (white) circles denote spins S=1 (s=1/2).}
\label{struct}
\end{figure}

\begin{figure}[t]
\begin{center}
\includegraphics[width=75mm]{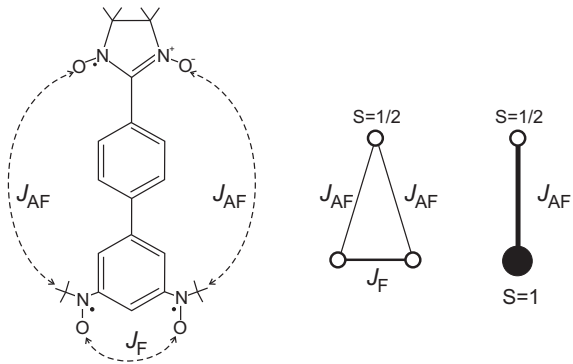}
\end{center}
\caption{Molecular structure of BIPNNBNO and the elementary magnetic  cell at $J_{\textrm{F}} \gg |J_{\textrm{AF}} |$.}
\label{molecule}
\end{figure}

 The crystal structure of BIPNNBNO is shown in Fig. \ref{struct}. A magnetic unit of the spin system presents organic triradical  BIPNNBNO.
Each of the molecules includes three s=1/2 spins (see Fig. \ref{molecule}) with intramolecular ferromagnetic $J_{F}$ and antiferromagnetic $J_{AF}$ interactions.  The magnitude of $|J_F| \sim 300$ K is very large, and two spins coupled ferromagnetically behave as a S - 1 moiety.    Ferrimagnetic chains are stretched along the b-axis. There are two kinds of antiferromagnetic interchain interactions along the a-axis. One is between the s-1/2 spins, which connects the nearest neighboring chains. The other is between S-1 species, which connects the next nearest neighboring chains and introduces spin frustration.

The puzzle is following. It is well known that a low-energy physics of an isolated ferrimagnetic (S,s) chain corresponds to a gapless (S-s) ferromagnet \cite{Brehmer1997}. Quite predictably one might expect an appearance of an  ordered state with fluctuations in the form of spin waves near the classical state. However, measurements of magnetization shows that an opening of a gap by analogy with the Haldane chain is likely scenario. Such a behavior is a manifestation of {\it non-Lieb-Mattis type ferrimagnetism} \cite{Sakai2011}. Namely, the magnetization measured at 400 mK is nearly zero below 4.5 T, increases rapidly above 4.5 T and exhibits a broad 1/3 plateau and a narrow 2/3 one at 7-23 T and around 26 T, respectively. Above 29 T, the magnetization is completely saturated \cite{Goto2003}.  

The purpose of the paper is to investigate the magnetization process. The problem is complicated by a lack of reliable  information about intra- and interchain  exchange interactions. So, before studying of  the 2D ferrimagnetic system BIPNNBNO we develop a simple quantum mechanical approach that  models a  magnetization process of the ferrimagnetic chain (1,1/2). The treatment agrees qualitatively with a predictions of the theory for quantum spin chains \cite{Oshikawa1997} and provides reasonable estimations of the exchange {\it intrachain} couplings.  In addition, it captures a peculiarity of a  magnetization process  in the  prototype 2D material, i.e. an appearance of the intermediate  2/3  plateau. Given these estimations we examine the magnetization process in BIPNNBNO by analyzing  exact diagonalization (ED)  calculations for a finite clusters of size $N=32$ and $N=40$. A main conclusion to be drawn from these calculations that an emergence of the anomalous singlet plateau  is a consequence of the frustrating interchain interaction. 

Two different mechanisms of formation of the plateau may be likely candidates: a generalization of Haldane’s conjecture to the weakly coupled ferrimagnetic chains,  and a valence-bond (dimerized) type ground state in the strong-coupling limit. To determine what of these scenarios is relevant  we develop low-energy effective theories for the 4-legs spin tube, which forms a minimal setup including the interchain couplings. In the regime of weakly coupled spin tube legs we apply abelian bosonization technique.    The opposite limit of a strong-ring interaction is analyzed   in terms of an effective  Heisenberg XXZ model where the intrachain coupling is perturbatively taken into account. Our analytical treatment shows  that only the first approach confirms an important role of frustration in stabilization of the singlet phase.

Note that a study of spin tubes is of interest by  itself because both of frustration and quantum fluctuation
are  strong \cite{Sakai2010}. Our model is directly related with the compound BIPNNBNO, but the main results are expected to apply to other frustrated spin tubes as well. Recently, it has been reported that the experimental candidate for the four-leg spin tube, Cu${}_2$Cl${}_4$$\cdot$D${}_8$C${}_4$SO${}_2$, is available \cite{Zheludev2008}.

The paper is organized as follows: In Sec. II we consider a magnetization process in the ferrimagnetic chain ($1,1/2$).  In Sec. III we discuss results of magnetization process in two-dimensional ferrimagnetic system obtained via the exact diagonalization method on a finite cluster. In Sec. IV we derive effective low-energy spin-1/2 Hamiltonian. A bosonisation study of the spin tube is carried out in Sec. V.  A discussion of these results is relegated to the Conclusion part.

\section{Magnetization of an isolated ferrimagnetic chain}

The issue that we address below is whether a calculation for an isolated ferrimagnetic chain partially reproduces  features of the magnetization curve observed in the BIPNNBNO crystal.  We demonstrate that both the 1/3 plateau and the  2/3 plateau can be recovered within a simple quantum mechanical analysis of a magnetization process of  an isolated  quantum $(1,1/2)$ ferrimagnetic  chain under an applied magnetic field. This enables to estimate the  intrachain exchange parameters $J_{AF}$ and $J_1$.

\begin{figure}[ht]
\begin{center}
\includegraphics[width=115mm]{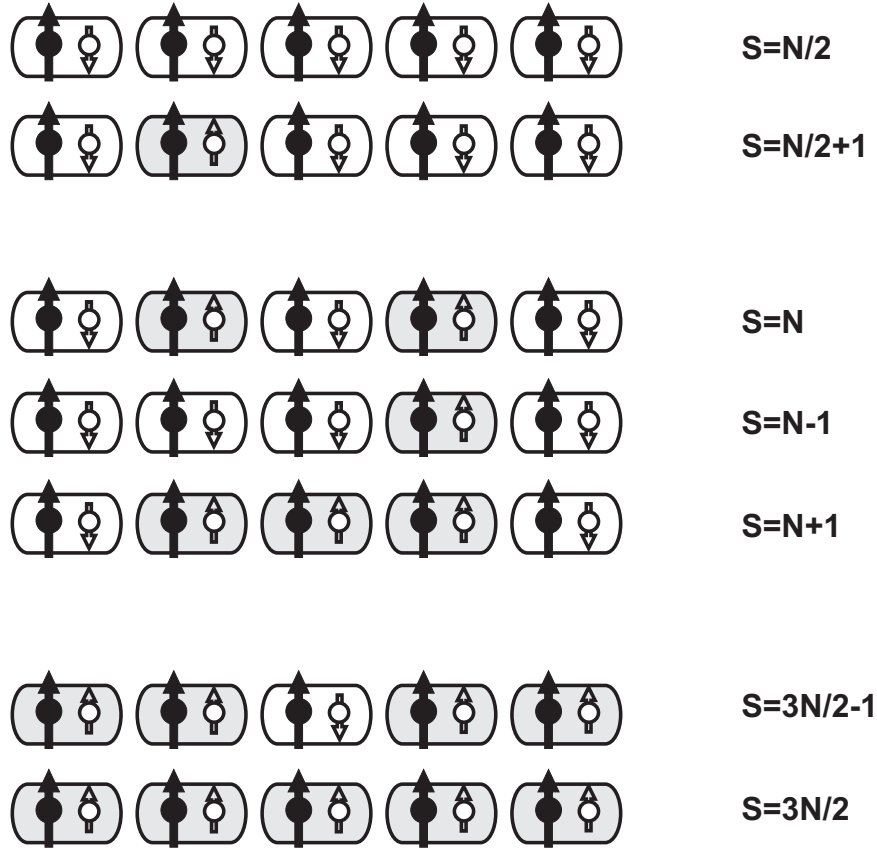}
\end{center}
\caption{Schematic picture of states of the ferrimagnetic chain used in construction of a magnetization curve. Excited blocks are marked by the gray shadow.}
\label{arrows}
\end{figure} 

We start from the values of the critical fields derived from the ED data \cite{Sakai1991} 
\begin{equation}
\left\{
\begin{array}{ccc}
B_1 & = & \left.  \frac{\partial \varepsilon}{\partial m} \right|_{m \to 1/3+0} \approx  E(N/2+1) - E(N/2), \\
B_2 & = & \left.  \frac{\partial \varepsilon}{\partial m} \right|_{m \to 2/3-0} \approx  E(N) - E(N-1),      \\
B_3 & = & \left.  \frac{\partial \varepsilon}{\partial m} \right|_{m \to 2/3+0} \approx  E(N+1) - E(N),     \\
B_{\textrm{sat}} & = & \left.  \frac{\partial \varepsilon}{\partial m} \right|_{m \to 1-0} \approx  E(3N/2) - E(3N/2-1),
\end{array}
\right.
\label{CritB}
\end{equation}
where $\varepsilon$, $m$ are the energy and the magnetization per an elementary cell of the $(1,1/2)$ ferrimagnetic chain, $N$ is a number of the elementary cells.

  To estimate the energies $E(S)$ in the right-hand side of Eqs.(\ref{CritB}), where $S$ is a total spin of the chain, we use the Hamiltonian of the one-dimensional quantum  ($1$,$1/2$)  ferrimagnet
\begin{equation}
\mathcal{\hat{H}}_{\textrm{c}} = J_{AF} \sum\limits_{i=1}^N \vec{S}_{1i}\vec{s}_{2i} + J_1 \sum\limits_{i=1}^N \vec{s}_{2i}\vec{S%
}_{1i+1}\;(S_{1}=1,\;s_{2}=1/2),  
\label{Hferri}
\end{equation}  
and construct the required quantum states $|S M \rangle$ with the given  quantum numbers of the total spin $S$ and its $z$-projection $M$.  

We suppose that the $1/3$ magnetization plateau corresponds to the ground state of the ($1$,$1/2$) ferrimagnetic chain with $S=N/2$.   The  wave function of the polarized state  is given by
\begin{equation}
| N/2,N/2 \rangle =  \prod_{i=1}^N  \left| \left(1 \frac12 \right) \frac12 \frac12 \right\rangle_i,
\label{gs}
\end{equation}
and presents a direct product of the spin states of the  magnetic elementary cells (1,1/2) [see Fig. \ref{arrows}].
The corresponding energy eigenvalue equals to 
\begin{equation}
 E(N/2)  = -J_{AF}N - \frac19 J_1 N.
\end{equation}

With an increasing of a magnetic field the ground state (\ref{gs}) is destroying and the state with $S=N/2+1$ stabilizes. A low-lying excitation may be qualitatively considered as a forming of one triplet bond.  The trial new wave function is
\begin{equation}
| N/2 +1,N/2+1 \rangle =  \frac{1}{\sqrt{N}} \sum_{k=1}^N  \left[ \prod_{i(\not= k)=1}^N  \left| \left(1 \frac12 \right) \frac12 \frac12 \right\rangle_i \right]   \left| \left(1 \frac12 \right) \frac32 \frac32 \right\rangle_k = \sum_{k=1}^N \alpha_k \Psi_k.
\end{equation} 
It is composed from all arrangements of the excited block within the chain taken with equal weights $\alpha_k$. 

By introducing the state and calculating the matrix element
\begin{equation}
\langle \Psi_k | \mathcal{\hat{H}}_{\textrm{c}} | \Psi_{k'} \rangle = \left[ E(N/2) +\frac32 J_{AF} +\frac{7}{18} J_1 \right] \delta_{kk^{'}} -\frac13 J_1 \delta_{k,k'\pm1} 
\end{equation}
one obtains the relationship for the coefficients $\alpha_k$  
\begin{equation}
-\frac13 J_1 \alpha_{k-1} + \left[ E(N/2) +\frac32 J_{AF} +\frac{7}{18} J_1 - E(N/2+1) \right] \alpha_k - \frac13 J_1 \alpha_{k+1} = 0, 
\end{equation}
which is tantamount to
\begin{equation}
E(N/2+1) = E(N/2) +\frac32 J_{AF} +\frac{7}{18} J_1 - \frac13 J_1 \left( \frac{\alpha_{k-1}}{\alpha_k}+\frac{\alpha_{k+1}}{\alpha_k} \right).
\end{equation}
This expression includes two independent variational parameters $\alpha_{k-1}/{\alpha_k}$ and $\alpha_{k+1}/{\alpha_k}$.  The minimal value 
\begin{equation}
E_{\textrm{min}}(N/2+1) = E(N/2) + \frac32 J_{AF} - \frac{5}{18} J_1
\end{equation}
is reached provided $\alpha_{k-1}/{\alpha_k}=\alpha_{k+1}/{\alpha_k}=1$. This yields the critical magnetic field $B_1$ destroying the $1/3$ plateau
\begin{equation}
B_1 = \frac32 J_{AF} - \frac{5}{18} J_1.
\label{B3}
\end{equation}

To find the critical fields $B_2$ and $B_3$ of the beginning and the end of the 2/3 plateau, respectively, we construct the trial states 
\begin{equation}
| N, N \rangle = \prod_{i=1}^{N/2}  \left| \left(1 \frac12 \right) \frac12 \frac12 \right\rangle_{2i-1} \left| \left(1 \frac12 \right) \frac32 \frac32 \right\rangle_{2i},
\end{equation}
$$
| N - 1,N - 1 \rangle =  \frac{1}{\sqrt{N}} \sum_{k=1}^{N/2}  \left[ \prod_{i=1}^{k-1}  \left| \left(1 \frac12 \right) \frac12 \frac12 \right\rangle_{2i-1} \left| \left(1 \frac12 \right) \frac32 \frac32 \right\rangle_{2i} \right]   
$$
\begin{equation}
\times \left| \left(1 \frac12 \right) \frac12 \frac12 \right\rangle_{2k-1} \left| \left(1 \frac12 \right) \frac12 \frac12 \right\rangle_{2k} \left[ \prod_{i=k+1}^{N/2}  \left| \left(1 \frac12 \right) \frac12 \frac12 \right\rangle_{2i-1} \left| \left(1 \frac12 \right) \frac32 \frac32 \right\rangle_{2i} \right],
\end{equation} 
$$
| N + 1,N + 1 \rangle =  \frac{1}{\sqrt{N}} \sum_{k=1}^{N/2}  \left[ \prod_{i=1}^{k-1}  \left| \left(1 \frac12 \right) \frac12 \frac12 \right\rangle_{2i-1} \left| \left(1 \frac12 \right) \frac32 \frac32 \right\rangle_{2i} \right]   
$$
\begin{equation}
\times \left| \left(1 \frac12 \right) \frac32 \frac32 \right\rangle_{2k-1} \left| \left(1 \frac12 \right) \frac32 \frac32 \right\rangle_{2k} \left[ \prod_{i=k+1}^{N/2}  \left| \left(1 \frac12 \right) \frac12 \frac12 \right\rangle_{2i-1} \left| \left(1 \frac12 \right) \frac32 \frac32 \right\rangle_{2i} \right],
\end{equation} 
which are schematically shown in Fig. \ref{arrows}. 

By the same manner we obtain
\begin{equation}
B_2 = \frac32 J_{AF} + \frac{7}{18} J_1, \quad B_3 = \frac32 J_{AF} + \frac{5}{6} J_1.
\label{B3B4}
\end{equation}

The saturation field $B_{\textrm{sat}}$ is determined with an aid of the trial wave functions
\begin{equation}
| 3N/2,3N/2 \rangle = \prod_{i=1}^N  \left| \left(1 \frac12 \right) \frac32 \frac32 \right\rangle_i,
\end{equation}
\begin{equation}
| 3N/2-1,3N/2-1 \rangle =  \frac{1}{\sqrt{N}} \sum_{k=1}^N  \left[ \prod_{i(\not= k)=1}^N  \left| \left(1 \frac12 \right) \frac32 \frac32 \right\rangle_i \right]   \left| \left(1 \frac12 \right) \frac12 \frac12 \right\rangle_k  
\end{equation} 
that results in
\begin{equation}
B_{\textrm{sat}} = \frac32 J_{AF} + \frac32 J_1.
\label{satur}
\end{equation}

\begin{figure}[ht]
\begin{center}
\includegraphics[width=115mm]{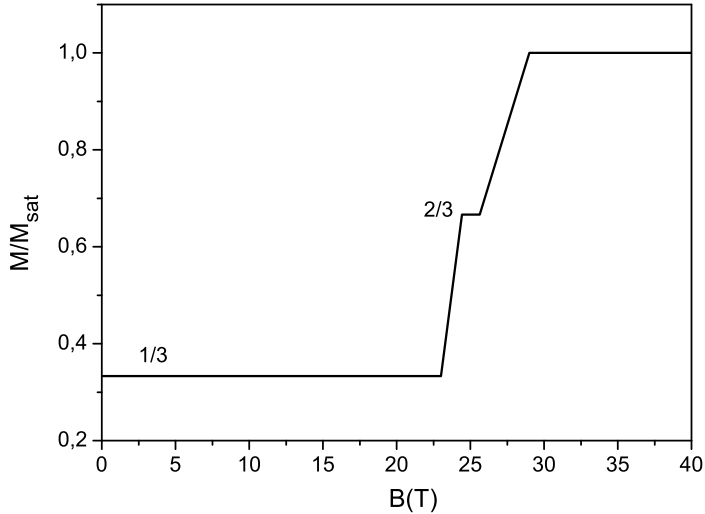}
\end{center}
\caption{Qualitative magnetization curve of the ferrimagnetic $(1,1/2)$ chain.}
\label{ferrimag}
\end{figure}

Given the experimental estimations for the 2D BIPNNBNO system, $B_1 \sim 31 \, K$, $B_2 \approx B_3 \sim 35 \, K$, and $B_{\textrm{sat}} = 39 \, K$,  we obtain from Eqs.(\ref{B3},\ref{satur}) the values of the intrachain exchange couplings, $J_{AF} \approx 21 \, K$ and $J_1 \approx 3.5 \, K$. By substituting them into Eq.(\ref{B3B4}) we get the critical fields of the 2/3 plateau, $B_2 \approx 32.8 \, K$ ($24.4 \, T$) and $B_3 \approx 34.4 \, K$ ($25.6 \, T$). A qualitative behavior of the ferrimagnetic chain  magnetization curve built from these reference points is depicted in Fig. \ref{ferrimag}.  We emphasize especially that an emergence of the intermediate 2/3 plateau is not related with interchain frustration effects. 

\section{Magnetization: Exact Diagonalization}
\begin{figure}[ht]
\includegraphics[width=110mm]{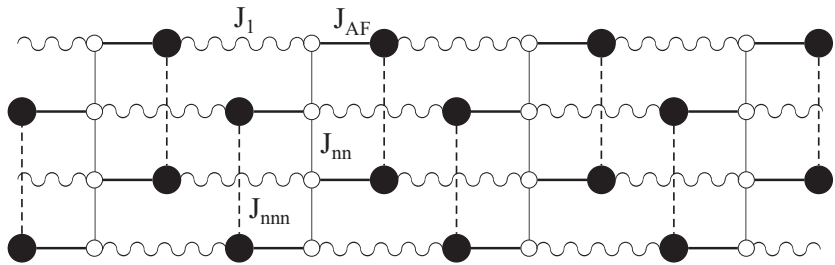}
\caption{Cluster of $N=32$ sites used in the exact diagonalization. The fixed exchange couplings are $J_{AF}=21 K$, $J_{1}=3.5 K$, $J_{\textrm{nn}} = 0.5 J_1 = 1.75 K$.}
\label{fig:cluster}
\end{figure}

In order to understand a role of the interchain couplings the magnetization process of the BIPNNBNO ferrimagnet was examined by the variant of a numerical diagonalization method with conservation of a total cluster spin \cite{Sinitsyn2007,Bostrem2010}.

The model Hamiltonian is given by  
\begin{equation}
\mathcal{\hat{H}}_{\textrm{clust}} =  J_{AF} \sum_{ij} {\vec S}_i {\vec s}_j + J_1 \sum_{ij} {\vec S}_i {\vec s}_j  + J_{\textrm{nn}} \sum_{ij}  {\vec s}_i {\vec s}_j + J_{\textrm{nnn}} \sum_{ij}  {\vec S}_i {\vec S}_j, 
\end{equation}
where $\vec{S}_i$ ($\vec{s}_i$) denotes spin-1 (spin-1/2) operator at site $i$.  The sublattices and the network of the antiferromagnetic interactions, $J_{AF}$, $J_1$, $J_{\textrm{nn}}$ and $J_{\textrm{nnn}}$, are shown in Fig. \ref{fig:cluster}.   We perform calculation of the N-step magnetization curve for the N=32 cluster depicted in the same Figure. The intrachain parameters, $J_{AF}$ and $J_1$, have been estimated in the previous Section whereas the interchain ones, $J_{\textrm{nn}}$ and $J_{\textrm{nnn}}$,   are assumed to be less than $J_1$. The open boundary conditions are used for the numerical calculations.

The magnetization process is compared with the results of the model of non-interacting (1,1/2) ferrimagnetic chains.  A standard way to build magnetization curve at $T=0$ is to define the lowest energy $E(N,M)$ of the Hamiltonian (\ref{Hferri}) in the subspace where $\sum_{j=1}^N \left( S^z_j + s^z_j \right) = M$ for a finite system of $N$ elementary $(S,s)$ blocks. Applying a magnetic field $B$ leads to a Zeeman splitting of the energy levels, and therefore level crossing occurs  on increasing the field. These level crossing correspond to jumps in the magnetization until the fully polarized state is reached at a certain value of the magnetic field. The magnetization of four independent chains  is then derived from
\begin{equation}
   m= 4 M/N, \quad M = \textrm{max} \left[ M | E(N,M+1) - E(N,M) > B \right], 
\end{equation}
which gives a step curve.

\begin{figure}
\includegraphics[width=120mm]{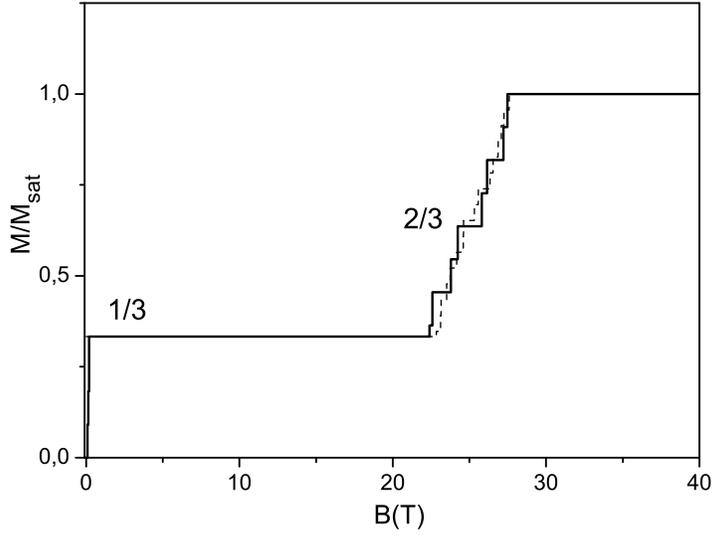}
\caption{Magnetization curve for the 32-site cluster. The exchange couplings are taken as $J_{AF}=21 K$, $J_{1}=3.5 K$, $J_{\textrm{nn}} = 0.5 J_1$, and $J_{\textrm{nnn}}=0$ (no frustration). The dotted line marks a calculation via the model of non-interacting $(1,1/2)$ chains.}
\label{fig:nstep1}
\end{figure}

\begin{figure}
\includegraphics[width=120mm]{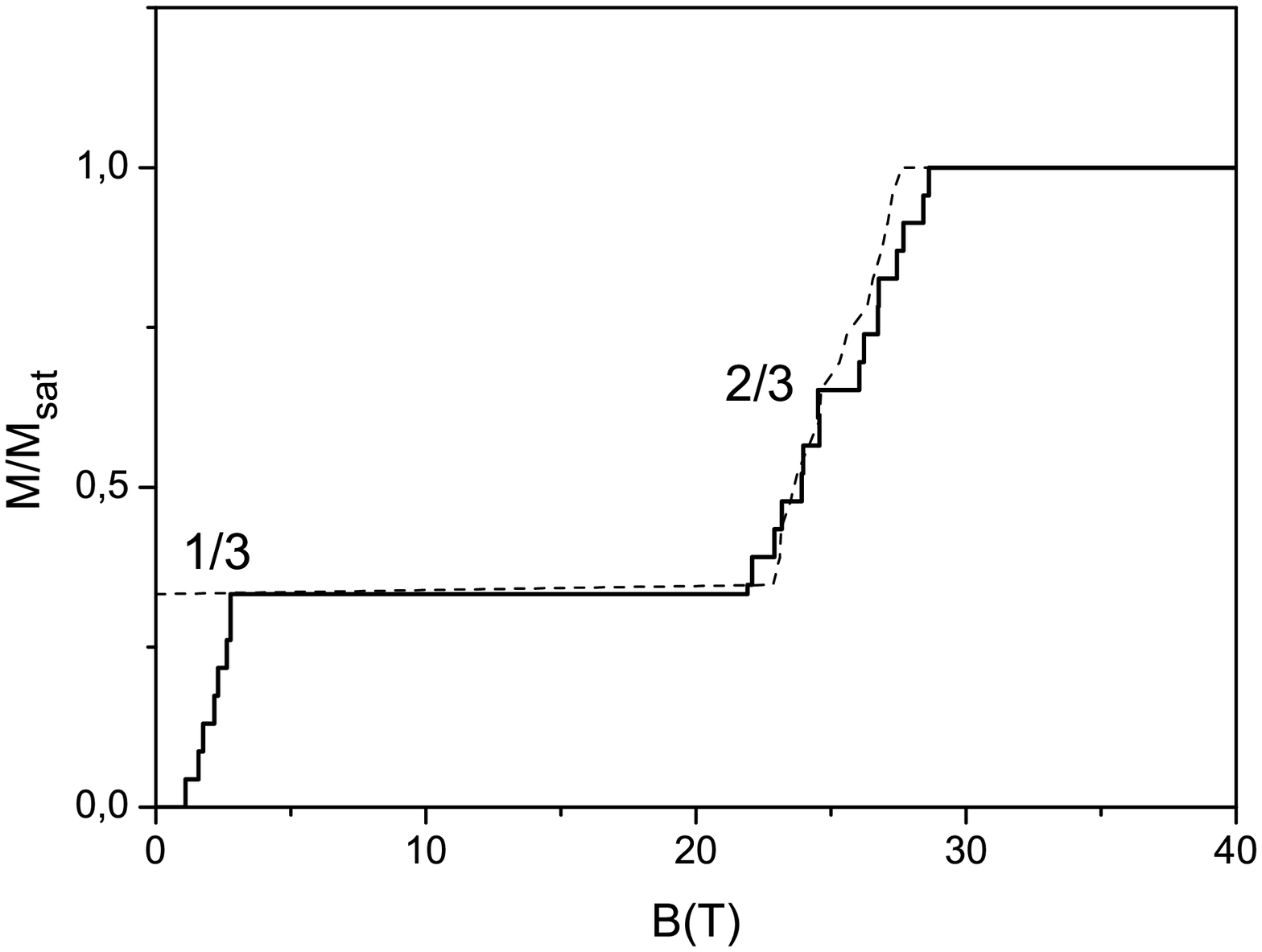}
\caption{Magnetization curve for the 32-site cluster. The exchange couplings are taken as $J_{AF}=21 K$, $J_{1}=3.5 K$, $J_{\textrm{nn}} = 0.5 J_1$, $J_{\textrm{nnn}}=0.075 J_1$. The dotted line marks a calculation via the model of non-interacting $(1,1/2)$ chains.}
\label{fig:nstep2}
\end{figure}

\begin{figure}
\includegraphics[width=120mm]{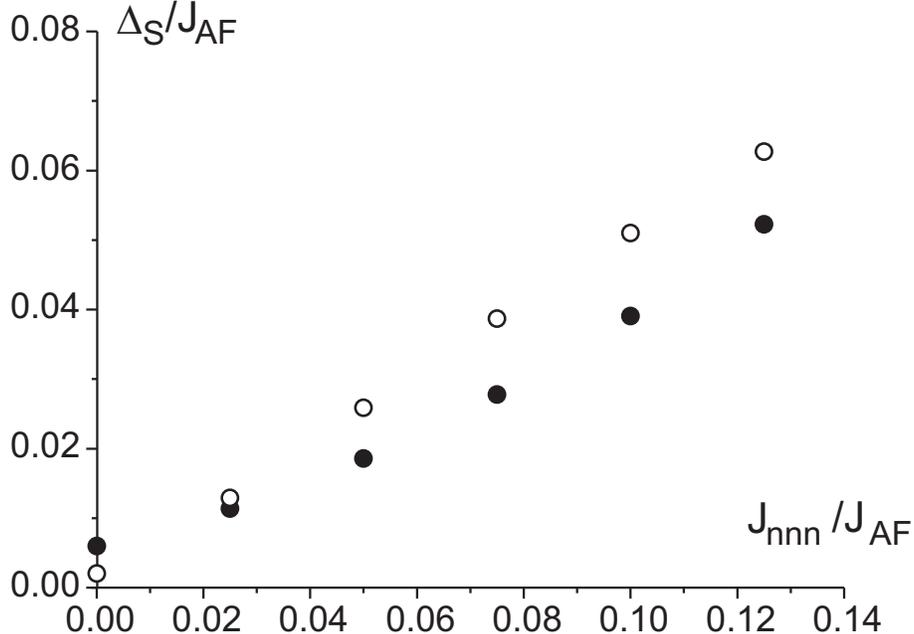}
\caption{Value of the singlet plateau $\Delta_S$ as a function of the frustrating coupling $J_{\textrm{nnn}}$ obtained on a cluster of size $N=32$ (black circles) and $N=40$ (white circles).}
\label{fig:Gap}
\end{figure}

An importance of the frustrating coupling is seen from comparison of two magnetization curves displayed in Figs. \ref{fig:nstep1}, \ref{fig:nstep2}. They correspond to no frustration case and  a  pronounced frustrating coupling, respectively.  The magnetization curves exhibit  several interesting features. For instance, the magnetization behavior in higher magnetic fields ($B>B_1$) is well reproduced within the model of non-interacting ferrimagnetic chains. Another remarkable feature revealed by  Figs. \ref{fig:nstep1}, \ref{fig:nstep2} that the singlet ground state plateau emerges at non-zero frustration interaction whereas the narrow $2/3$ plateau appears regardless of  the frustration.  We numerically found that the  width $\Delta_S$  of the singlet  plateau scales almost linearly with a $J_{\textrm{nnn}}$ value (Fig. \ref{fig:Gap}). To check into the case of the dependence we repeat cacluations  on a cluster of larger size, $N=40$, with the same set of parameters that support the finding. The observation points out that the zero magnetization plateau has a quantum origin with a crucial role of frustration which destroys a long-range order and drives the system into the singlet phase. Below, we address analytically the issue  in the regimes of strong and weak interchain couplings.

\section{A formation of the singlet plateau}

In low-dimensional Heisenberg systems frustrating couplings can drive transitions to gapfull quantum states, where local singlets form a ground state. These quantum gapped phases may have long-ranged singlet order (valence bond state), or realize a resonating valence  band spin liquid. In last case, a ground state is a coherent  superposition of all lattice-coverings by local singlets \cite{Misguich2005}.
 
\begin{figure}
\includegraphics[width=120mm]{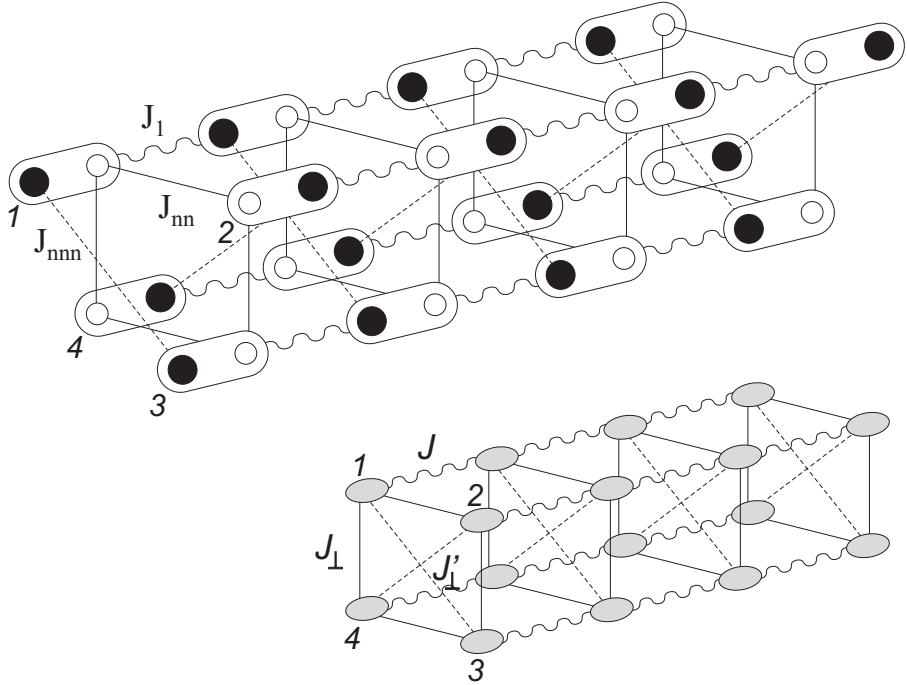}
\caption{4-leg spin tube structure used in the limit of strong ring coupling (up) and in the case of weakly interacting chains (below). The black (white) circles mean spin-1 (spin-1/2) sites. The gray circles denote renormalized spin-1/2 blocks.}
\label{fig:TubeRing}
\end{figure}

To recognize features of these phases in the ED results  we undertake analytical treatments of the four-legs spin tube shown in Fig. \ref{fig:TubeRing}. The new system is infinite  along the $b$-axis, and periodic with the 4-site period along the $a$-axis.  The tube forms a minimal setup including the interchain nearest- and next-to-nearest neighbor  couplings and contains the same number of ferrimagnetic chains parallel to the b-axis  as the clusters in the ED study. As we demonstrate below, the simplified  model elucidates an  important role of frustration in stabilization of the singlet phase.

The singlet phase may arise in the limit of strong ring coupling $J_{\textrm{nn}}$, $J_{\textrm{nnn}} \gg J_1$. In this case,   the problem can be analyzed in terms of  Heisenberg XXZ model similar to ladders in a magnetic field \cite{Mila1998}.  The opposite limit ($J_1 \gg J_{\textrm{nn}}$, $J_{\textrm{nnn}}$) results in a scenario of weakly interacting chains. Based on a block renormalization procedure the original system is then mapped onto  the model of a spin tube with four ferromagnetic spin-1/2 legs. We mention that the ground state properties of two-leg spin ladders  with ferromagnetic intrachain coupling and antiferromagnetic interchain couplings have been discussed in Refs. \cite{Kolezhuk1996,Roji1996} in an absence of an external field. A magnetization process of those spin ladders with an even number of legs (2 and 4) has been studied in Ref. \cite{Wiessner1999} in the regime of weak ferromagnetic coupling along the legs and strong antiferromagnetic coupling along the rungs.

 An appearance of the singlet phase in the frustrated spin tube with four weakly coupled ferromagnetic spin-1/2 legs can be studied through the bosonization technique which proves effectiveness for quasi-one-dimensional spin-one-half systems. To the best of our knowledge, the system has never been previously reported, however our further analysis follows closely to that of given in Ref. \cite{Kim1999}, where 4-legs spin tube with antiferromagnetic chains and a specific form of diagonal rung interactions (but with no frustration) has been treated.  Note as well that spin ladders with ferromagnetic and  ferrimagnetic legs are much less studied \cite{Japaridze2007,Aristov2004} by the bosonization approach  in comparison with ladder systems with  antiferromagnetic legs. The main problem arising here is that the  formalism is well defined only if there is an easy-plane exchange anisotropy. In this regard, we note that measurements of the angular dependence of the ESR linewidth for the BIPNNBNO system showed that the largest linewidth was observed for the field direction perpendicular to the ab plane \cite{Kanzawa2006}. Due to the theoretical consideration by Oshikawa and Affleck\cite{Oshikawa2002}  a critical regime of XY-anisotropy is expected  in the compound.  In addition we point out that the ED algorithm invoked in the previous Section enables to treat clusters of sufficiently large sizes due to use of the rotational $SU(2)$ symmetry. The latter is broken  by the anisotropy whose role in a singlet gap formation is a  subject of future ED studies.


\subsection{Spin tube: weakly interacting rings and a    model of a single XXZ chain}

We study the Hamiltonian of the spin tube (see Fig. \ref{fig:TubeRing})
\begin{equation}
\mathcal{H} =  \sum_{n=1}^N H^{\textrm{ring}}_{n}
+ J_1 \sum_{n=1}^N \left( 
\textbf{S}_{n,1}  \textbf{s}_{n+1,1} + 
\textbf{s}_{n,2}  \textbf{S}_{n+1,2} +
\textbf{S}_{n,3}  \textbf{s}_{n+1,3}+
\textbf{s}_{n,4}  \textbf{S}_{n+1,4}
\right)
- B \sum_{n=1}^N \sum_{i=1}^4 \left( S^z_{n,i} + s^z_{n,i} \right), 
\label{HRing} 
\end{equation}
where the Hamiltonian of the separate ring is
$$
H^{\textrm{ring}}_{n} = J_{\textrm{nn}} \left( 
  \textbf{s}_{n,1}\textbf{s}_{n,2} +
  \textbf{s}_{n,2}\textbf{s}_{n,3} +
  \textbf{s}_{n,3}\textbf{s}_{n,4} +
  \textbf{s}_{n,4}\textbf{s}_{n,1}  
\right)  
+ J_{\textrm{nnn}} \sum_{n=1}^N \left( 
  \textbf{S}_{n,1}\textbf{S}_{n,3} +
  \textbf{S}_{n,2}\textbf{S}_{n,4} 
\right). 
$$
Here, $S=1$ and $s=1/2$, $n$ is the index of the ring, $N$ is the total number of rings,  and the index $i$ marks the (1,1/2) blocks inside the rings. Periodic boundary conditions along the tube direction are imposed. In our model it is suggested that  $J_{\textrm{nn}}$, $J_{\textrm{nnn}} \gg J_1$. 

In the limit $J_1 = 0$ the system decouples into a collection of nonineracting rings. At zero magnetic field the  singlet and triplet states
$$
| \psi_0 \rangle = - \frac{\sqrt{3}}{2} | 00;00 \rangle +\frac12 | 11;00 \rangle,
$$
\begin{equation}
| \psi_1 \rangle =  \frac{1}{\sqrt{2}} | 01;11 \rangle  + \frac{1}{\sqrt{2}}  | 10;11 \rangle
\label{PSIring}
\end{equation}
have the lowest energies $E_0 = - 2 J_{\textrm{nn}}/9 + 8 J_{\textrm{nnn}}/9$ and $E_1 = - J_{\textrm{nn}}/9 + 8 J_{\textrm{nnn}}/9$, respectively. The states of the ring $| S_{12}S_{34};SM \rangle$ are obtained via the common rule of addition of moments, where $S_{12}$ ($S_{34}$) is spin of dimer composed of the spins of the 1 and 2 (3 and 4) blocks.  The singlet and triplet states of the ring that enter into (\ref{PSIring}) are given in Appendix A.

Upon increasing the magnetic field a transition between the singlet and triplet states occurs at  $B=J_{\textrm{nn}}/9$ and the the total magnetization jumps abruptly from zero to $M=N$. 

At non-zero ring coupling the sharp transition is broadened and starts from a critical value $B_0$. 
To find the field we  derive  the XXZ  spin chain Hamiltonian by  we using  the standard approach that is analogous to study of spin-1/2 ladder with strong rung exchange \cite{Mila1998}.

The Hamiltonian (\ref{HRing}) is splitted into two parts
$$
     \mathcal{H} = \mathcal{H}_0 + \mathcal{H}_1.
$$
$$
\mathcal{H}_0 = \sum_{n=1}^N H^{\textrm{ring}}_{n} - B_c \sum_{i=1}^4 \sum_{n=1}^N \left( S^z_{i n} + s^z_{i n} \right),
$$
$$
\mathcal{H}_1 = J_1 \sum_{n=1}^N \left[ 
\textbf{S}_{n,1}  \textbf{s}_{n+1,1} + 
\textbf{s}_{n,2}  \textbf{S}_{n+1,2} +
\textbf{S}_{n,3}  \textbf{s}_{n+1,3}+
\textbf{s}_{n,4}  \textbf{S}_{n+1,4}
\right]
 -
\left( B-B_c \right) \sum_{i=1}^4 \sum_{n=1}^N ( S^z_{i n} + s^z_{i n}),
$$ 
where $B_c = E_1 - E_0$.  
The $\mathcal{H}_1$ lifts the $2^N$-fold degeneracy of the ground state of the Hamiltonian  $\mathcal{H}_0$ . The later can be either in the state $| \psi_0 \rangle$ or $| \psi_1 \rangle$.  By using the standard many body perturbation theory \cite{Fulde} the effective Hamiltonian can be derived
\begin{equation}
  \mathcal{H}_{\textrm{eff}} = J^{\textrm{eff}}_{xy} \sum_{n=1}^N \left( \tilde{S}^x_n \tilde{S}^x_{n+1} + \tilde{S}^y_n  \tilde{S}^y_{n+1} \right) + J^{\textrm{eff}}_{z} \sum_{n=1}^N  \tilde{S}^z_n \tilde{S}^z_{n+1}  - H^{\textrm{eff}}_z  \sum_{n=1}^N  \tilde{S}^z_n ,
  \label{HeffS}
\end{equation}
where $J^{\textrm{eff}}_{xy} = -16 J_1/27$, $J^{\textrm{eff}}_{z} = - J_1/9$ and  $B^{\textrm{eff}}_z = J_1/9 + B-B_c$.

To get the expression the pseudo-spin $\tilde{S}_i =1/2$  operators that act on the states $| \psi_0 \rangle$ and $| \psi_1  \rangle$ are introduced
$$
 \tilde{S}^z_n | \psi_0 \rangle_n = -\frac12  | \psi_0 \rangle_n, \qquad \tilde{S}^z_n | \psi_1 \rangle_n = \frac12  | \psi_1 \rangle_n,
$$
$$
 \tilde{S}^{+}_n | \psi_0 \rangle_n =  | \psi_1 \rangle_n, \qquad \tilde{S}^{+}_n | \psi_1 \rangle_n = 0,
$$
\begin{equation}
\tilde{S}^{-}_n | \psi_0 \rangle_n = 0, \qquad  \tilde{S}^{-}_n | \psi_1 \rangle_n =  | \psi_0 \rangle_n .
\label{PseudoSpin}
\end{equation}

The starting spin-1 operators and the pseudo-spin operators in the restricted space are related by
$$
   S^z_{i n}   = \frac{1}{6} + \frac{1}{3} {\tilde S}^z_{n},
$$
\begin{equation}
   S^{+}_{i n} =  (-1)^{i-1} \frac{4}{3\sqrt{3}}  {\tilde S}^{+}_{n}, \qquad 
   S^{-}_{i n} =  (-1)^{i-1} \frac{4}{3\sqrt{3}}  {\tilde S}^{-}_{n}. 
\end{equation}
The corresponding map for the spin-1/2 operators is 
$$
    s^z_{i n} = -\frac{1}{24} - \frac{1}{12} {\tilde S}^z_{n},
$$
\begin{equation}
    s^{+}_{i n} = \frac{(-1)^i}{3\sqrt{3}} {\tilde S}^{+}_{n}, \qquad 
    s^{-}_{i n} = \frac{(-1)^i}{3\sqrt{3}} {\tilde S}^{-}_{n}.
\end{equation}

The Jordan-Wigner transformation maps the Hamiltonian (\ref{HeffS}) onto a system of interacting spinless fermions
\begin{equation}
    \mathcal{H}_{sf} = t \sum_{n=1}^N \left( c^{+}_i c_{i+1} + c^{+}_{i+1} c_i \right) + V \sum_{n=1}^N n_i n_{i+1} - \mu \sum_{n=1}^N n_i,
\end{equation}
where $t  = J^{\textrm{eff}}_{xy}/2$, $V = J^{\textrm{eff}}_{z}$ and $\mu = J^{\textrm{eff}}_{z} + B^{\textrm{eff}}_z$. 

The lowest critical field $B_0$ corresponds to that value of the chemical potential $\mu$ for which
the band of spinless fermions starts to fill up. This yields the condition $\mu = -2 t$ and leads to the result 
\begin{equation}
    B_0 = \frac19 J_{\textrm{nn}} + \frac{16}{27} J_{1}.
\end{equation} 
The critical value  involves no  frustration  parameter $J_{\textrm{nnn}}$ that is clearly contrary to  the ED  results.  

A similar analysis can be carried out for the saturation field. The details of the calculations are relegated to Appendix B.

\subsection{Spin tube: a model of weakly interacting   ferromagnetic legs and abelian bosonization}

To apply the bosonization  we should map the initial system, consisting of two sorts of spins,   spin-1/2 and spin-1, to the spin-1/2 system by using the  quantum renormalization group (QRG) in real space based on the block renormalization procedure \cite{pfeuty}. To exploit the real-space QRG technique one divide the spin lattice into small blocks, namely, the intrachain dimers (1,1/2), and obtains the
lowest energy states $\left\{ \left| \alpha \right\rangle \right\} $ of each  isolated block. The effect of inter-block interactions is then taken into account by constructing an effective Hamiltonian $H_{\textrm{eff}}$ which now acts on
a smaller Hilbert space embedded in the original one. In this new Hilbert space each of the former blocks is treated as a single site.  The effective  Hamiltonians $H_{\textrm{eff}}=Q^{\dagger }HQ$ is constructed via the projection operator $Q=\prod\limits_{i=1}^{N}Q_{i}$ with $Q_{i}=\sum\limits_{\alpha=1}^{m}\left| \alpha \right\rangle \left\langle \alpha \right| $ of each $i$-th block where $m$ is the number of low energy states that are kept and $N$
is a number of lattice cells.

We hold the lowest doublet $S-1/2$ to find an effective low-energy Hamiltonian. The higher energy $S-3/2$ states are neglected. One can check that the reduced matrix elements are ($S_1=1$, $s_2=1/2$)
$$
\langle 1 \, 1/2; 1/2 || S_1 || 1 \, 1/2; 1/2  \rangle = 2 \sqrt{\frac23},
$$
$$
\langle  1 \, 1/2;1/2||s_2 ||1 \, 1/2;1/2> = - \frac{1}{\sqrt{6}}.
$$
Therefore the effective spin-1/2 operators of the renormalized chain are 
\begin{equation}
Q_{i}^{\dagger }\vec{S}_{1i}Q_{i}=\frac{4}{3}\vec{S}_{i},\;Q_{i}^{\dagger }%
\vec{s}_{2i}Q_{i}=-\frac{1}{3}\vec{S}_{i} \qquad (S=1/2).  
\label{QRspin}
\end{equation}
The renormalized  Hamiltonian of the intrachain interactions corresponds to the ferromagnetic Heisenberg spin-1/2 model with the exchange coupling $J=-4 J_1/9$ (Fig. \ref{fig:TubeRing}). The interchain interactions between the nearest neighbors, spins -1/2, and next to the nearest neighbors, spins -1, are renormalized as $J_{\perp}=J_{\textrm{nn}}/9$ and $J^{'}_{\perp}=16 J_{\textrm{nnn}}/9$, respectively.

Consider a four-legs spin tube consisting of spin-$1/2$ chains. The Hamiltonian of the system is
\begin{equation}
\mathcal{{\hat H}}_{\textrm{tube}} = \sum_{\lambda=1}^4 \mathcal{{\hat H}}_{\lambda} + \mathcal{{\hat H}}^{\perp}_{12}+\mathcal{{\hat H}}^{\perp}_{23}+\mathcal{{\hat H}}^{\perp}_{34}+\mathcal{{\hat H}}^{\perp}_{14}+\mathcal{{\hat H}}^{'\,\perp}_{13}+\mathcal{{\hat H}}^{'\,\perp}_{24}.
\label{HamiltonTube}
\end{equation}
The spins along the chains are coupled ferromagnetically, the Hamiltonian  for the separate $\lambda$-th chain is
$$
  \mathcal{{\hat H}}_{\lambda} =  - J^{xy} \sum_{i=1}^N \left( S^x_{\lambda,j}S^x_{\lambda,j+1} + S^y_{\lambda,j}S^y_{\lambda,j+1} \right) - J^z \sum_{i=1}^N S^z_{\lambda,j}S^z_{\lambda,j+1},
$$  
where $S^{x,y,z}_{\lambda,j}$ are the spin S=1/2 operators at the $j$th site, the intraleg coupling is ferromagnetic, $J>0$.

The interaction parts are given by 
\begin{equation}
\mathcal{{\hat H}}^{\perp}_{\lambda \lambda'} = J^{xy}_{\perp,\lambda \lambda'} \sum_{j=1}^N \left( S^x_{\lambda,j}S^x_{\lambda',j} + S^y_{\lambda,j}S^y_{\lambda',j} \right) + J^z_{\perp,\lambda \lambda'} \sum_{i=1}^N S^z_{\lambda,j}S^z_{\lambda,j},
\label{NNint}
\end{equation}
\begin{equation}
\mathcal{{\hat H}}^{'\,\perp}_{\lambda \lambda'} =  
J^{'\,xy}_{\perp,\lambda \lambda'} \sum_{j=1}^N \left( S^x_{\lambda,j}S^x_{\lambda',j} + S^y_{\lambda,j}S^y_{\lambda',j} \right) + J^{'\,z}_{\perp,\lambda \lambda'} \sum_{i=1}^N S^z_{\lambda,j}S^z_{\lambda,j}
\label{NNNint}
\end{equation}
and includes the nearest, $J_{\perp}>0$, and the next-to-nearest, $J^{'}_{\perp}>0$, antiferromagnetic interleg couplings.

The unitary transformation keeping spin commutation relations
$$
S^{x,y}_{\lambda,j} \to (-1)^j S^{x,y}_{\lambda,j}, \quad S^{z}_{\lambda,j} \to S^{z}_{\lambda,j}
$$
maps the Hamiltonian (\ref{HamiltonTube}) to the Hamiltonian with antiferromagnetic legs. It changes $J^{xy} \to -J^{xy}$ and  $J^z \to J^z$, and  the ferromagnetic isotropic point is $\Delta= J^z/J^{xy}=-1$ in the Hamiltonian with the antiferromagnetic legs.

Following the general procedure of transforming a spin model to an effective model of continuum field, we convert the spin Hamiltonian of the spin tube with {\it antiferromagnetic} legs to a Hamiltonian of spinless fermions using Jordan-Wigner transformation, then map it to a modified Luttinger model. The bosonic expressions for spin operators are
$$
S^{+}_{\lambda} (x) = \frac{S^{+}_{j \lambda}}{a} = \frac{e^{- i \sqrt{\pi} \Theta_{\lambda}}}{\sqrt{2\pi a}} \left[ e^{-  i (\pi x/a) } + \cos \left( \sqrt{4\pi} \Phi_{\lambda} \right) \right],
$$
\begin{equation}
S^{z}_{\lambda} (x) = \frac{S^z_{j \lambda}}{a} =  \frac{1}{\sqrt{\pi}} \partial_x \Phi_{\lambda} + \frac{1}{\pi a} e^{i (\pi x/a) } \sin \left(\sqrt{4\pi} \Phi_{\lambda} \right),
\end{equation}
where $\Phi$ and $\Theta$ are the bosonic dual fields, and  $x$ is defined on the lattice, $x_j= j a$, $a$ is a short-distance cutoff.

The bosonized form of the Hamiltonian of the non-interacting chains is
\begin{equation}
\mathcal{H}_{\lambda} =  \frac{u}{2} \int dx \, \left[ K \Pi^2_{\lambda} + \frac{1}{K} \left( \partial_x \Phi_{\lambda} \right)^2 \right], 
\end{equation}
where $\Pi_{\lambda}(x)=\partial_x \Theta_{\lambda}$ is canonically conjugate momentum to $\Phi_{\lambda}$. The Luttinger liquid parameters are fixed from the Bethe ansatz solution \cite{Luther1975}
\begin{equation}
K = \frac{\pi}{2(\pi - \arccos \Delta)}, \qquad u = J^{xy} \, \frac{\pi \sqrt{1-\Delta^2}}{2 \arccos \Delta}.
\end{equation}
The velocity $u$ vanishes and $K$ diverges for $\Delta=-1$. This corresponds to the ferromagnetic instability point of a single chain. 

The interchain interactions (\ref{NNint}) between the nearest neighbor chains  reads as 
$$
\mathcal{H}^{\perp}_{\lambda \lambda'} = J^z_{\perp,\lambda \lambda^{'}} \int \frac{dx}{\pi} \left( \partial_x \Phi_{\lambda} \right) \left( \partial_x \Phi_{\lambda^{'}} \right)  + g_1 \int \frac{d x}{(2\pi a)^2} \cos \left( \sqrt{4 \pi} (\Phi_{\lambda}+\Phi_{\lambda^{'}}) \right)
$$
$$
+ g_2 \int \frac{d x}{(2\pi a)^2} \cos \left( \sqrt{4 \pi} (\Phi_{\lambda}-\Phi_{\lambda^{'}}) \right)
+ g_3 \int \frac{d x}{(2\pi a)^2} \cos \left( \sqrt{\pi} (\Theta_{\lambda}-\Theta_{\lambda^{'}}) \right) 
$$
$$
+ g_4 \int \frac{d x}{(2\pi a)^2} \cos \left( \sqrt{\pi} (\Theta_{\lambda}-\Theta_{\lambda^{'}}) \right) \cos \left( \sqrt{4 \pi} (\Phi_{\lambda}+\Phi_{\lambda^{'}}) \right) 
$$
\begin{equation}
+ g_5 \int \frac{d x}{(2\pi a)^2} \cos \left( \sqrt{\pi} (\Theta_{\lambda}-\Theta_{\lambda^{'}}) \right) \cos \left( \sqrt{4 \pi} (\Phi_{\lambda}-\Phi_{\lambda^{'}}) \right), 
\end{equation}
where $g_1  = -2 J^z_{\perp,\lambda \lambda^{'}}$, $g_2  = 2 J^z_{\perp,\lambda \lambda^{'}}$, $g_3  = 2 \pi J^{xy}_{\perp,\lambda \lambda^{'}}$, $g_4 = g_5 = \pi J^{xy}_{\perp,\lambda \lambda^{'}}$.  The Hamiltonian (\ref{NNNint}) of the next-to-nearest couplings has a similar form with a formal change $J^z_{\perp,\lambda \lambda^{'}} \to J^{'\,z}_{\perp,\lambda \lambda^{'}}$, $g_1 \to g^{'}_{1} = -2 J^{'\,z}_{\perp,\lambda \lambda^{'}}$, $g_2 \to g^{'}_{2} = 2 J^{'\,z}_{\perp,\lambda \lambda^{'}}$ etc.

Following the route of Refs.\cite{Kim1999,Cabra1998} it is convenient to introduce a symmetric mode $\Phi_s$  and three antisymmetric ones $\Phi_{a_1}$, $\Phi_{a_2}$, $\Phi_{a_3}$ 
\begin{equation}
\begin{array}{c}
\Phi_s = \frac12 \left( \Phi_1 + \Phi_2 +  \Phi_3 + \Phi_4 \right), \\
\Phi_{a_1} = \frac12 \left( \Phi_1 + \Phi_2 -  \Phi_3 - \Phi_4 \right), \\
\Phi_{a_2} = \frac12 \left( \Phi_1 - \Phi_2 -  \Phi_3 + \Phi_4 \right), \\
\Phi_{a_3} = \frac12 \left( \Phi_1 - \Phi_2 +  \Phi_3 - \Phi_4 \right).
\end{array}
\end{equation}

In terms of the new fields the quadratic part of the Hamiltonian (\ref{HamiltonTube}) is diagonalized to 
\begin{equation}
\mathcal{H}_0 = \frac{u_s}{2} \int dx \, \left[ K_s \Pi^2_s + \frac{1}{K_s} \left(\partial_x \Phi_s \right)^2 \right] +
\sum_{i=1}^3 \frac{u_{a_i}}{2} \int dx \,  \left[ K_{a_i} \Pi^2_{a_i} + \frac{1}{K_{a_i}} \left(\partial_x \Phi_{a_i} \right)^2 \right]
\label{Hzero}
\end{equation}
with 
$$
u_s = u \left( 1+ \frac{2K J^z_{\perp}}{u\pi} + \frac{K J^{'\,z}_{\perp}}{u \pi} \right)^{\frac12}, \qquad 
K_s = K \left( 1+ \frac{2K J^z_{\perp}}{u\pi} + \frac{K J^{'\,z}_{\perp}}{u \pi} \right)^{-\frac12},
$$
$$
u_{a_1} = u_{a_2} =  u \left( 1- \frac{K J^{'\,z}_{\perp}}{u \pi} \right)^{\frac12}, \qquad 
K_{a_1} = K_{a_2} =  K \left( 1- \frac{K J^{'\,z}_{\perp}}{u \pi} \right)^{-\frac12},
$$
\begin{equation}
u_{a_3} = u \left( 1- \frac{2K J^z_{\perp}}{u\pi} + \frac{K J^{'\,z}_{\perp}}{u \pi} \right)^{\frac12}, \qquad 
K_{a_3} = K \left( 1- \frac{2K J^z_{\perp}}{u\pi} + \frac{K J^{'\,z}_{\perp}}{u \pi} \right)^{-\frac12}.
\end{equation}

The relevant and marginally relevant terms of the interchain couplings are given by
$$ 
\mathcal{H}_{\textrm{int}} = 2 g_1 \sum_{i=1}^2 \int \frac{d x}{(2\pi a)^2} \cos \left( \sqrt{4\pi} \Phi_s \right) \cos \left( \sqrt{4\pi} \Phi_{a_i}\right) + 
2 g^{'}_1  \int \frac{d x}{(2\pi a)^2} \cos \left( \sqrt{4\pi} \Phi_s \right) \cos \left( \sqrt{4\pi} \Phi_{a_3}\right)
$$
$$
+ 2 g_2 \sum_{i=1}^2 \int \frac{d x}{(2\pi a)^2} \cos \left( \sqrt{4\pi} \Phi_{a_i} \right) \cos \left( \sqrt{4\pi} \Phi_{a_3}\right) + 
2 g^{'}_2  \int \frac{d x}{(2\pi a)^2} \cos \left( \sqrt{4\pi} \Phi_{a_1} \right) \cos \left( \sqrt{4\pi} \Phi_{a_2}\right)
$$
$$
+ 2 g_3 \sum_{i=1}^2 \int \frac{d x}{(2\pi a)^2} \cos \left( \sqrt{\pi} \Theta_{a_i} \right) \cos \left( \sqrt{\pi} \Theta_{a_3}\right) + 
2 g^{'}_3  \int \frac{d x}{(2\pi a)^2} \cos \left( \sqrt{\pi} \Theta_{a_1} \right) \cos \left( \sqrt{\pi} \Theta_{a_2}\right)
$$ 
$$
+ g_4  \int \frac{d x}{(2\pi a)^2} \left[ 
\cos \left( \sqrt{\pi} (\Theta_{a_2}+\Theta_{a_3}) \right) \cos \left( \sqrt{4\pi} (\Phi_s+\Phi_{a_1}) \right)
\right.
$$
$$
+ \cos \left( \sqrt{\pi} (\Theta_{a_1}-\Theta_{a_3}) \right) \cos \left( \sqrt{4\pi} (\Phi_s-\Phi_{a_2}) \right)
+ \cos \left( \sqrt{\pi} (\Theta_{a_2}-\Theta_{a_3}) \right) \cos \left( \sqrt{4\pi} (\Phi_s-\Phi_{a_1}) \right)
$$
$$
\left. 
+ \cos \left( \sqrt{\pi} (\Theta_{a_1}+\Theta_{a_3}) \right) \cos \left( \sqrt{4\pi} (\Phi_s+\Phi_{a_2}) \right)
\right]
$$
$$
+ g^{'}_4  \int \frac{d x}{(2\pi a)^2} \left[ 
\cos \left( \sqrt{\pi} (\Theta_{a_1}+\Theta_{a_2}) \right) \cos \left( \sqrt{4\pi} (\Phi_s+\Phi_{a_3}) \right)
\right.
$$
\begin{equation}
\left. 
+ \cos \left( \sqrt{\pi} (\Theta_{a_1}-\Theta_{a_2}) \right) \cos \left( \sqrt{4\pi} (\Phi_s-\Phi_{a_3}) \right)
\right] .
\label{Hint}
\end{equation}
The $g_5$ terms are irrelevant and are omitted.

The Hamiltonian (\ref{Hzero}) describes four independent gapless spin-$1/2$ chains coupled by the interchain interaction in the form of Eq.(\ref{Hint}). It is expected that  the interleg coupling results in the Haldane gap in the excitation spectrum. Note that in the vicinity of the single chain ferromagnetic instability, $\Delta = -1$, the effective bandwidth collapses, $u \to 0$, and the effect of the interleg couplings becomes crucial.  To find a detail behavior of the gap in the phase diagram with antiferromagnetic ($J_{\perp}$,  $J^{'}_{\perp}>0$) interleg coupling and ferromagnetic leg regime ($\Delta < 0$) we use the renormalization group analysis. 

The RG equations are derived through the standard technique (see Ref.\cite{Giamarchi}, for example). The result is
$$
\frac{d g_1}{dl} = \left[2 - \left( K_s + K_{a_1} \right) \right] g_1,
$$
$$
\frac{d g^{'}_1}{dl} = \left[2 - \left( K_s + K_{a_3} \right) \right] g^{'}_1,
$$
$$
\frac{d g_2}{dl} = \left[2 - \left( K_{a_3} + K_{a_1} \right) \right] g_2,
$$
$$
\frac{d g^{'}_2}{dl} = \left[2 - 2  K_{a_1} \right] g^{'}_2,
$$
$$
\frac{d g_3}{dl} = \left[2 - \frac14 \left( \frac{1}{K_{a_1}} + \frac{1}{K_{a_3}} \right) \right] g_3,
$$
$$
\frac{d g^{'}_3}{dl} = \left[2 -   \frac{1}{2 K_{a_1}} \right] g^{'}_3,
$$
$$
\frac{d g_4}{dl} = \left[2 - \left( K_s + K_{a_1} + \frac{1}{4 K_{a_1}} + \frac{1}{4 K_{a_3}} \right) \right] g_4,
$$
$$
\frac{d g^{'}_4}{dl} = \left[2 - \left( K_s + K_{a_3} + \frac{1}{2 K_{a_1}} \right) \right] g^{'}_4,
$$
$$
\frac{d K_s}{dl} = - 4 g^2_1 \left( \frac{K_s}{4 \pi u_s} \right)^2 - 2 {g^{'}_1}^2 \left( \frac{K_s}{4 \pi u_s} \right)^2 - 2 g^2_4 \left( \frac{K_s}{4 \pi u_s} \right)^2 -  {g^{'}_4}^2 \left( \frac{K_s}{4 \pi u_s} \right)^2,
$$
$$
\frac{d K_{a_1}}{dl} = - \frac12 g^2_1 \left( \frac{K_{a_1}}{2 \pi u_{a_1}} \right)^2 - \frac12 g^2_2 \left( \frac{K_{a_1}}{2 \pi u_{a_1}} \right)^2 -  \frac12 {g^{'}_2}^2 \left( \frac{K_{a_1}}{2 \pi u_{a_1}} \right)^2 - \frac14 g^2_4 \left( \frac{K_{a_1}}{2 \pi u_{a_1}} \right)^2
$$
$$
+\frac12 \left( \frac{g_3}{4 \pi u_{a_1}} \right)^2 + \frac12 \left( \frac{g^{'}_3}{4 \pi u_{a_1}} \right)^2
+\frac14 \left( \frac{g_4}{4 \pi u_{a_1}} \right)^2 + \frac14 \left( \frac{g^{'}_4}{4 \pi u_{a_1}} \right)^2,
$$
$$
\frac{d K_{a_3}}{dl} = - \frac12 {g^{'}_1}^2 \left( \frac{K_{a_3}}{2 \pi u_{a_3}} \right)^2 -  g^2_2 \left( \frac{K_{a_3}}{2 \pi u_{a_3}} \right)^2  - \frac14 {g^{'}_4}^2 \left( \frac{K_{a_3}}{2 \pi u_{a_3}} \right)^2
$$
\begin{equation}
+ \left( \frac{g_3}{4 \pi u_{a_3}} \right)^2 +\frac12 \left( \frac{g_4}{4 \pi u_{a_3}} \right)^2 .
\label{RGeq}
\end{equation}
One sees that the $g_1$ terms are relevant for $K_s + K_{a_1} <2$; the $g^{'}_1$ term is relevant for $K_s + K_{a_3}<2$; the $g_2$ terms are relevant for $K_{a_3} + K_{a_1} <2$; the $g^{'}_2$ term is relevant for $K_{a_1}<1$; the $g_3$ term is relevant for $ K^{-1}_{a_1} + K^{-1}_{a_3} <8$; the $g^{'}_3$ term is relevant for $K_{a_1}>1/4$. Despite the $g_4$ and $g^{'}_4$ terms are irrelevant they are the most relevant terms which couple  the symmetric and antisymmetric modes \cite{Kim1999}.  

Using the RG equations the behavior of the gap in the whole phase diagram can be established. Following to standard routine, we analyze the effect of the transversal ($J^{xy}_{\perp}$)  and the longitudinal ($J^{z}_{\perp}$)  parts of the interleg coupling separately.

\subsubsection{Transversal part of the interleg interactions}

In this case $J^{xy}_{\perp},J^{'\,xy}_{\perp}\neq 0$ and $J^{z}_{\perp},J^{'\,z}_{\perp} = 0$, the initial values of the coupling constants are given by $g_1(l=0) = g^{'}_1(l=0) = 0$, $g_2(l=0) = g^{'}_2(l=0) = 0$,  $g_3(l=0)=2 \pi J^{xy}_{\perp}$,  $g^{'}_3(l=0) =2 \pi J^{'\,xy}_{\perp}$, $g_4(l=0)=\pi J^{xy}_{\perp}$ and $g^{'}_4(l=0) =\pi J^{'\,xy}_{\perp}$. The bare Luttinger parameters are $u_s=u$, $u_{a_1}=u_{a_2}=u_{a_3}=u$, and $K_s(l=0)=K$, $K_{a_1}(l=0)=K_{a_2}(l=0)=K_{a_3}(l=0)=K$. The term $g_3$ is relevant for $-1 \leq \Delta \leq 0$ while the $g_4$ term is irrelevant. It is easily checked numerically that $g_3$, $g^{'}_3$ grow whereas $g_4$, $g^{'}_4$ decrease under the RG. It means that $\Theta_1$, $\Theta_2$, $\Theta_3$ are locked in one of the vacuum states ($\Theta_1=\Theta_2=0$, $\Theta_3=\sqrt{\pi}$ or $\Theta_1=\Theta_2=\sqrt{\pi}$, $\Theta_3=0$ provided $J^{xy}_{\perp}>$$J^{'\,xy}_{\perp}$), fluctuations of the fields $\Theta_{a_i}$ ($i=1,2,3$) are completely suppressed.  Therefore arbitrary $J^{xy}_{\perp}>J^{'\,xy}_{\perp} \neq 0$  generate a gap in the antisymmetric modes ($\Theta_i$ are pinned, disordered).

After the fluctuations of the fields $\Theta_i$ are stopped, the infrared behavior of the symmetric mode is governed by the term of the coupling with the antisymmetric modes
$$
\mathcal{\tilde{H}}_{\textrm{int}} =   2 \tilde{g}_4 \sum_{i=1}^2 \int \frac{dx}{(2\pi a)^2} \cos (\sqrt{4\pi} \Phi_s)  \cos (\sqrt{4\pi} \Phi_{a_i}) 
$$
\begin{equation}
+  2 \tilde{g}^{'}_4 \int \frac{dx}{(2\pi a)^2} \cos (\sqrt{4\pi} \Phi_s)  \cos (\sqrt{4\pi} \Phi_{a_3}).
\end{equation}
where  
$$
  \tilde{g}_4 = \bar{g}_4 \langle \cos \left( \sqrt{\pi} [\Theta_{a_1}+\Theta_{a_3}] \right) \rangle = \bar{g}_4 \langle \cos \left( \sqrt{\pi} [\Theta_{a_2}+\Theta_{a_3}] \right) \rangle ,
$$
$$
  \tilde{g}^{'}_4 = \bar{g}^{'}_4 \langle \cos \left( \sqrt{\pi} (\Theta_{a_1}+\Theta_{a_2}) \right) \rangle ,
$$
and $\bar{g}_4$, $\bar{g}^{'}_4$ are renormalized couplings provided $g_3,g^{'}_3 = \mathcal{O}(1)$. Here, the invariance of $\mathcal{\tilde{H}}_{\textrm{int}}$ given by Eq.(\ref{Hint}) under $\Theta_{a_i} \to - \Theta_{a_i}$ yields $\langle \cos \left( \sqrt{\pi} (\Theta_{a_i}-\Theta_{a_j}) \right) \rangle = \langle \cos \left( \sqrt{\pi} (\Theta_{a_i}+\Theta_{a_j}) \right) \rangle$. 
Despite $e^{i\Phi_{a_i}}$ has exponentially decaying correlations due to the $\Theta_{a_i}$ are pinned, a scrupulous analysis \cite{Kim1999} shows that the effective Hamiltonian for $\Phi_s$ presents a standard sine-Gordon Hamiltonian 
\begin{equation}
\mathcal{\tilde{H}}_{\textrm{eff}} = \frac{u_s}{2} \int dx \, \left[ \bar{K}_s \Pi^2_s + \frac{1}{\bar{K}_s} \left(\partial_x \Phi_s \right)^2 \right] + g \int \frac{d x}{(2\pi a)^2} \cos \left( \sqrt{16 \pi} \Phi_s \right),
\end{equation}  
where $\bar{K}_s$ is a renormalized value of $K$, and $g$ is a new effective coupling constant. From the correlation function $\langle \exp \left[ i \sqrt{16\pi} \Phi_s (x) \right] \exp \left[ i \sqrt{16\pi} \Phi_s (y) \right] \rangle = \left( a^2/|x-y|^2 \right)^{4 \bar{K}_s}$ it follows that the $g$ term has a scale dimension $4 \bar{K}_s$. Therefore, it is relevant for $\bar{K}_s<1/2$, when $\Phi_s$ is pinned, i.e. becomes massive \cite{Tsvelik}.

\begin{figure}[ht]
\begin{center}
\includegraphics[width=100mm]{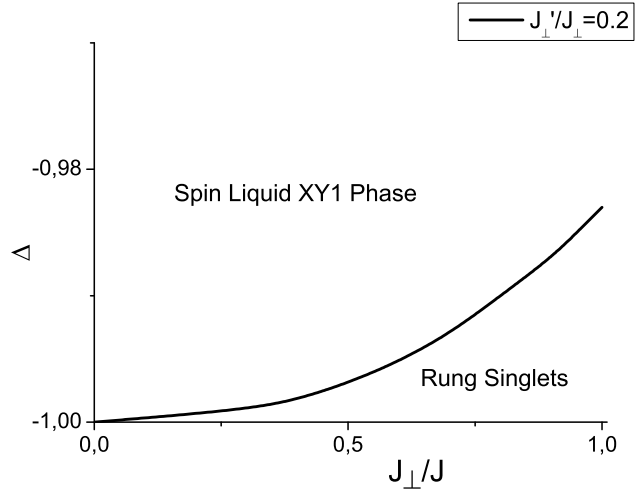}
\end{center}
\caption{The ground-state phase diagram in the vicinity $\Delta=-1$ of the four-leg tube with transverse coupling between legs.}%
\label{XYPD}
\end{figure}

To summarize, the transversal part of the interleg coupling supports gapped antisymmetric modes, the symmetric sector is gapped at $\bar{K}_s <1/2$, and remains gapless at $\bar{K}_s > 1/2$. The condition $\bar{K}_s = 1/2$ determines a  boundary between the gapless {\it Spin Liquid XY1 phase} \cite{Shulz}, and a generalization of the gapped {\it Rung Singlets} phase \cite{Vekua2003} for the for-leg spin tube (Fig.\ref{XYPD}). (Hereinafter, we retain names of phases used in the theory of spin ladders with ferromagnetic legs.) In the last case, spins on the same rungs, or along the shortest diagonals   form singlet pairs by a dynamical way.

\subsubsection{Longitudinal part of the interleg interactions}

For the case of the longitudinal part of the interleg exchange, $J^{xy}_{\perp},J^{'\,xy}_{\perp}=0$ and $J^{z}_{\perp},J^{'\,z}_{\perp}\neq 0$, the bare values of the coupling constants are given by $g_1(l=0) = -2 J^{z}_{\perp}$,   $g^{'}_1(l=0) = -2 J^{'\,z}_{\perp}$, $g_2(l=0) = 2 J^{z}_{\perp}$,   $g^{'}_2(l=0) = 2 J^{'\,z}_{\perp}$,   $g_3(l=0)=g^{'}_3(l=0) =0$, and $g_4(l=0)=g^{'}_4(l=0) =0$.

\begin{figure}[ht]
\begin{center}
\includegraphics[width=120mm]{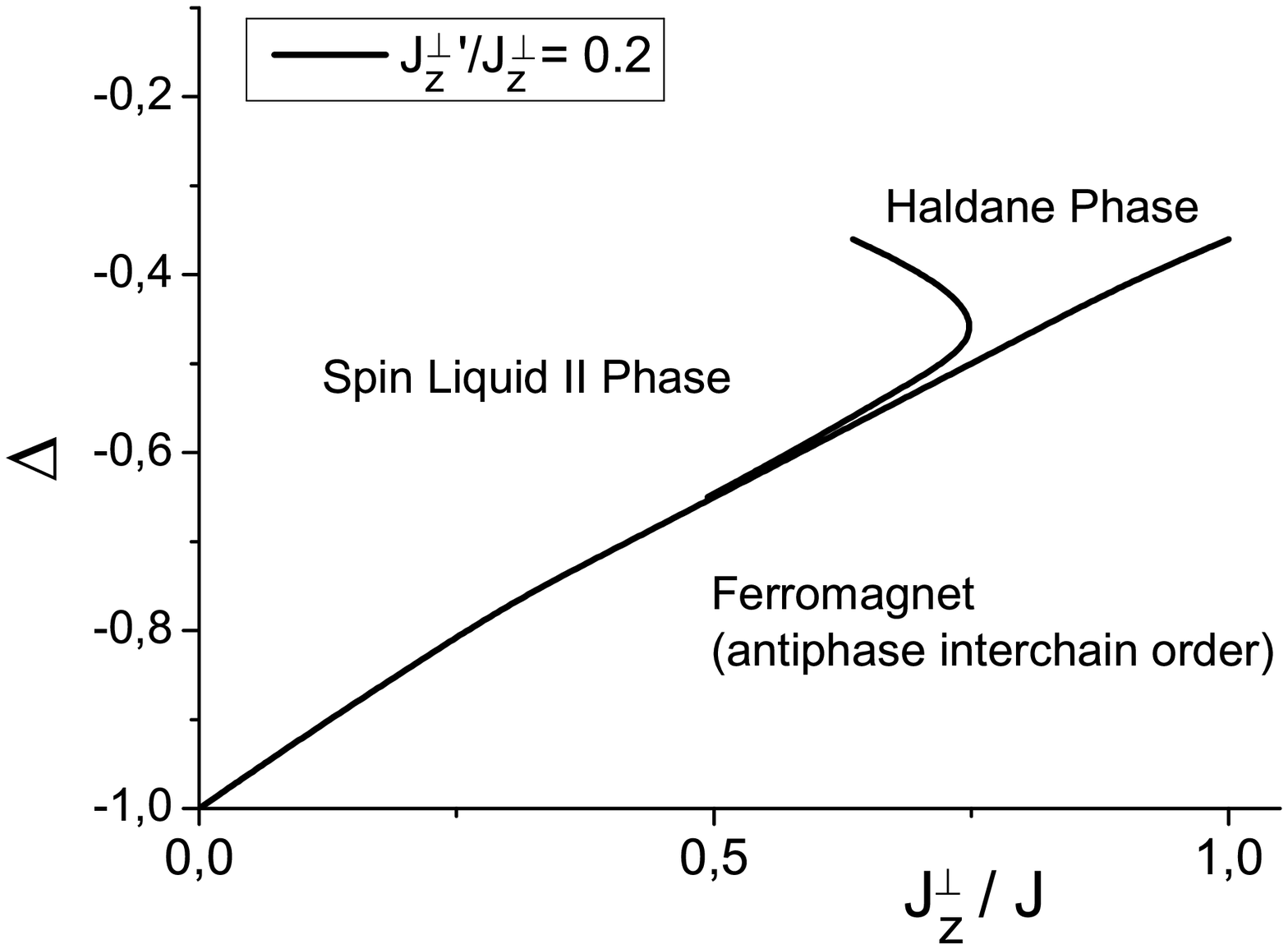}
\end{center}
\caption{The ground-state phase diagram in the vicinity $\Delta=-1$ of the four-leg tube with longitudinal coupling between legs.}%
\label{ZPD}
\end{figure}

The strong-coupling phase diagram in the vicinity of ferromagnetic instability point ($\Delta=-1$ and $J^{z}_{\perp},J^{'\,z}_{\perp} = 0$) obtained in the RG analysis is shown in Fig. \ref{ZPD}. In the sector denoted as {\it spin liquid II phase}\cite{Vekua2003} the $g_{1,2}$, $g^{'}_{1,2}$ terms  are irrelevant. The symmetric and antisymmetric modes remain gapless. In the sector marked as a {\it Haldane phase} the terms  $g_{1,2}$, $g^{'}_{1,2}$ are relevant. Since all of the modes are coupled and locked together, both the symmetric and antisymmetric modes are gapped.  
 
The phase of {\it a ferromagnet with antiphase interchain order} arises as a result of the ferromagnetic instability with increasing interleg antiferromagnetic coupling. The boundary of the transition into the phase is obtained by studying the velocity renormalization  of the corresponding gapless excitations. We mark the transition at $u_{a_i}=0$ ($i=1,2,3$).

\subsubsection{Isotropic interleg exchange}
The initial values of the coupling constants are $g_1(l=0) = -2 J_{\perp}$,   $g^{'}_1(l=0) = -2 J^{'}_{\perp}$, $g_2(l=0) = 2 J_{\perp}$,   $g^{'}_2(l=0) = 2 J^{'}_{\perp}$,  $g_3(l=0)=2 \pi J_{\perp}$,  $g^{'}_3(l=0) =2 \pi J^{'}_{\perp}$, $g_4(l=0)=\pi J_{\perp}$ and $g^{'}_4(l=0) =\pi J^{'}_{\perp}$.

\begin{figure}[ht]
\begin{center}
\includegraphics[width=120mm]{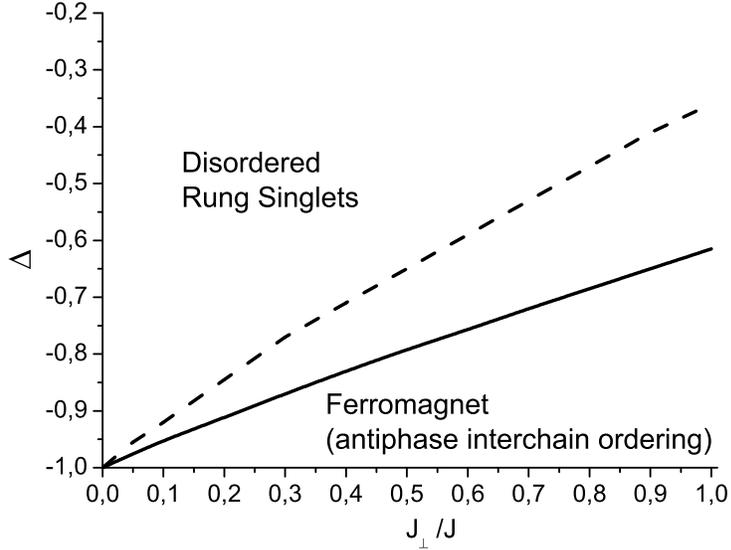}
\end{center}
\caption{The ground-state phase diagram in the vicinity $\Delta=-1$ of the four-leg tube with isotropic coupling between legs. The phase boundary between the disordered rung singlet gapfull phase and the stripe ferromagnetic phase is shown by the dotted and solid lines for $J^{'}_{\perp} / J_{\perp} =0.2$ and $J^{'}_{\perp} / J_{\perp} =1.0$, respectively.}%
\label{IZPD}
\end{figure}

From the RG equations (\ref{RGeq}) it is seen that the most relevant operators are the $g_3$, $g^{'}_3$ terms. Therefore, the antisymmetric sector is gapped, and $\Theta_{a_i}$ are locked in the disordered phase. As in the case of the transversal interleg interactions an effective sine-Gordon   Hamiltonian for the symmetric mode determines phase boundary $\bar{K}_s=1/2$ between gapped and gapless phases. Numerical analysis shows that the ground state phase diagram consists of the disordered Rung Singlet gapfull phase and the stripe ferromagnetic phase with dominating intraleg ferromagnetic ordering.  The sector of the rung singlet phase increases  at $J^{'}_{\perp} \to J_{\perp}$  (see Fig. \ref{IZPD}).

A possible physical picture that reconciles the phase diagram with the results of the previous sections might look as follows. The system BIPNNBNO is an  array  of {\it loosely} coupled ferrimagnetic chains in a presence of an {\it extremely} weak XY anisotropy and a strong frustration, $J^{'}_{\perp} \sim J_{\perp}$, what corresponds to the region of the disordered Rung Singlet phase close the point of the ferromagnetic instability, $\Delta = -1$.

\section{Conclusion}

We have studied magnetization process in two-dimensional compound BIPNNBNO which exhibits ferrimagnetism of    non-Lieb-Mattis type.  The investigation is complicated by a lack of reliable information about exchange interactions in the system. For a start, we proposed the naive model of non-interacting ferrimagnetic chains and showed that an appearance both 1/3 and 2/3 plateaus can be explained within the model. This provides us the intrachain exchange couplings $J_{AF}$ and $J_1$.  By setting these parameters in the exact diagonalization routine  the magnetization curve for the 32 and 40-sites clusters is numerically calculated. We demonstrate that the magnetization curve  similar to that observed in the experiment is obtained in the regime of weak interchain coupling, $J_1 \gg J_{\textrm{nn}} > J_{\textrm{nnn}}$.  Another revealed phenomenon is that a width of the singlet plateau increases with a growth of the antiferromagnetic  frustrating coupling between the next-to-nearest chains. Following these results, we apply on the tube lattice  two low-energy theories which could explain an appearance of the singlet phase.  The first one is based on the effective XXZ Heisenberg model in a longitudinal magnetic field in the limit where the interchain coupling dominates, $J_{\textrm{nn}} \geq J_{\textrm{nnn}} \gg J_1$. We derive the  critical field destroying the singlet plateau, it turns out that it does not depend on the frustration parameter  $J_{\textrm{nnn}}$. Another analytical strategy is realized via the abelian bosonization formalism which is relevant for the opposite limit   $J_1 \gg J_{\textrm{nn}} \geq J_{\textrm{nnn}}$. We demonstrate that the gapfull disordered {\it Rung Singlets} phase comes up when the XY exchange anisotropy may tilt the balance from the long-range order  with an  antiphase interchain arrangement  of ferrimagnetic chains towards the spin liquid phase. A  role of the anisotropy in a formation of the spin gap in the original two-dimensional system deserves a further study.

\begin{acknowledgments}
We acknowledge Grant RFBR No. 10-02-00098-a for a support. We wish to thank Prof. D.N. Aristov for discussions.
\end{acknowledgments}

\appendix

\section{Wave functions of the ring}

The states of the ring $|S_{12}S_{34}SM\rangle$  are composed from the functions $| m_1 m_2 m_3 m_4 \rangle$, where $m_i = \pm 1/2$ marks the spin-1/2 state of the separate i-th (1,1/2) block in the ring
$$
     \left| 1 \frac12; \frac12  \pm \frac12 \right\rangle = \pm  \sqrt{\frac23} \left| 1 \pm 1 \right\rangle \left| \frac12 \mp \frac12 \right\rangle \mp \frac{1}{\sqrt{3}} \left| 10 \right\rangle \left| \frac12  \pm \frac12 \right\rangle.
$$

The basic functions of the singlet states are given by
$$
| 00;00 \rangle = \frac12 \left| \frac12 -\frac12 \frac12 -\frac12 \right\rangle  
-\frac12 \left| \frac12 -\frac12  - \frac12 \frac12 \right\rangle
-\frac12 \left| -\frac12 \frac12  \frac12 -\frac12 \right\rangle 
+\frac12 \left| -\frac12 \frac12  - \frac12 \frac12 \right\rangle , 
$$
$$
| 11;00 \rangle = \frac{1}{\sqrt{3}} \left| \frac12 \frac12 -\frac12 -\frac12 \right\rangle
+  \frac{1}{\sqrt{3}} \left| -\frac12 -\frac12 \frac12 \frac12 \right\rangle
-\frac{1}{2\sqrt{3}} \left| \frac12 -\frac12 \frac12 -\frac12 \right\rangle
$$
$$
-\frac{1}{2\sqrt{3}} \left| \frac12 -\frac12 -\frac12 \frac12 \right\rangle
-\frac{1}{2\sqrt{3}} \left| -\frac12 \frac12 \frac12 -\frac12 \right\rangle
-\frac{1}{2\sqrt{3}} \left| -\frac12 \frac12 -\frac12 \frac12 \right\rangle .
$$

The triplet states read as 
$$
| 01;11 \rangle = \frac{1}{\sqrt{2}} \left| \frac12 -\frac12 \frac12 \frac12 \right\rangle
-  \frac{1}{\sqrt{2}} \left| -\frac12 \frac12 \frac12 \frac12 \right\rangle  ,
$$
$$
| 10;11 \rangle = \frac{1}{\sqrt{2}} \left| \frac12 \frac12 \frac12 -\frac12 \right\rangle
-  \frac{1}{\sqrt{2}} \left| \frac12 \frac12 -\frac12 \frac12 \right\rangle  .
$$

\section{Saturation field in the limit of the strong ring coupling}

To get the saturation field the functions of the ring with the total spins $S=5$ and $S=6$ 
$$
| \psi_6 \rangle = \left| \frac32 \frac32 \right\rangle_1 \left| \frac32 \frac32 \right\rangle_2 \left| \frac32 \frac32 \right\rangle_3 \left| \frac32 \frac32 \right\rangle_4, 
$$
$$
| \psi_5 \rangle = -\frac12 \left| \frac12 \frac12 \right\rangle_1 \left| \frac32 \frac32 \right\rangle_2 \left| \frac32 \frac32 \right\rangle_3 \left| \frac32 \frac32 \right\rangle_4 
+  \frac12 \left| \frac32 \frac32 \right\rangle_1 \left| \frac12 \frac12 \right\rangle_2 \left| \frac32 \frac32 \right\rangle_3 \left| \frac32 \frac32 \right\rangle_4
$$
$$
- \frac12 \left| \frac32 \frac32 \right\rangle_1 \left| \frac32 \frac32 \right\rangle_2 \left| \frac12 \frac12 \right\rangle_3 \left| \frac32 \frac32 \right\rangle_4 
+ 
\frac12 \left| \frac32 \frac32 \right\rangle_1 \left| \frac32 \frac32 \right\rangle_2 \left| \frac32 \frac32 \right\rangle_3 \left| \frac12 \frac12 \right\rangle_4. 
$$
with the energies $E_5 = J_{\textrm{AF}}/2  -J_{\textrm{nn}}/3 + J_{\textrm{nnn}}/2$, $E_6 = 2 J_{\textrm{AF}} +   J_{\textrm{nn}} + 2 J_{\textrm{nnn}}$ are needed.

By introducing the pseudo-spin operators in the restricted space similar to Eq.(\ref{PseudoSpin})   the original spin operators are presented as follows ($S=1$, $s=1/2$)
$$
   S^z_{i n}   = \frac{23}{24} + \frac{1}{12} {\tilde S}^z_{n}, \qquad 
   s^z_{i n}   = \frac{5}{12} + \frac{1}{6} {\tilde S}^z_{n},
$$
$$
   S^{+}_{i n} =  (-1)^{i+1} \frac12 \sqrt{\frac23}   {\tilde S}^{+}_{n}, \qquad
   s^{+}_{i n} =  (-1)^{i} \frac12 \sqrt{\frac23}   {\tilde S}^{+}_{n},
$$
\begin{equation}
   S^{-}_{i n} =  (-1)^{i+1} \frac12 \sqrt{\frac23}  {\tilde S}^{-}_{n}, \qquad
   s^{-}_{i n} =  (-1)^{i} \frac12 \sqrt{\frac23}  {\tilde S}^{-}_{n}.  
\end{equation}

The effective XXZ spin chain Hamiltonian has the form (\ref{HeffS}) with the parameters  $J^{\textrm{eff}}_{xy} = -2 J_1/3$,  $J^{\textrm{eff}}_{z} =  J_1/18$ and  $B^{\textrm{eff}}_z =  - 7 J_1/9 + B-B_c$, where $B_c = E_6 - E_5$.

By performing a Jordan-Wigner transformation followed by a particle-hole transformation \cite{Mila1998} the new Hamiltonian of spinless holes is
$$
\mathcal{H}_{h} = - t \sum_{n=1}^N \left( d^{+}_i d_{i+1} + d^{+}_{i+1} d_i \right) + V \sum_{n=1}^N n^d_i n^d_{i+1} - \mu_h \sum_{n=1}^N n^d_i,
$$
where $t  = J^{\textrm{eff}}_{xy}/2$, $V = J^{\textrm{eff}}_{z}$ and  
$\mu_h = J^{\textrm{eff}}_{z} -  B^{\textrm{eff}}_z$.

The saturation field $B_{\textrm{sat}}$ corresponds to the chemical potential where the hole band starts to fill up, $\mu_h = -2 t$. This yields 
$ B_{\textrm{sat}} = B_c  + J_1/6 = 3 J_{\textrm{AF}}/2 +   4 J_{\textrm{nn}}/3 + 3 J_{\textrm{nnn}}/2 + J_1/6$.

\end{document}